\RequirePackage{fix-cm}
\documentclass[smallextended]{svjour3}       
\smartqed  
\usepackage{graphicx}
\usepackage{nicefrac}
\usepackage{verbatim}
\usepackage{amsmath}
\usepackage{bm}
\usepackage{amssymb}
\usepackage{mathrsfs}

\newcommand{\eref}[1]{(\ref{#1})}
\newcommand{\eps }{\varepsilon }

\begin{document}

\title{Bound states of spin-half particles in a static gravitational field close to the black hole field}


\author{A.~F.~Spencer-Smith \and
    G.~H.~Gossel \and
    J.~C.~Berengut \and
    V.~V.~Flambaum
}


\institute{School of Physics, University of New South Wales, Sydney 2052, Australia  \\
\email{g.gossel@unsw.edu.au}
}

\date{Received: date / Accepted: date}

\maketitle

\begin{abstract}
We consider the bound-state energy levels of a spin-1/2 fermion in the gravitational field of a near-black hole object. In the limit that the metric of the body becomes singular, all binding energies tend to the rest-mass energy (i.e. total energy approaches zero). We present calculations of the ground state energy for three specific interior metrics (Florides, Soffel and Schwarzschild) for which the spectrum collapses and becomes quasi-continuous in the singular metric limit. The lack of zero or negative energy states prior to this limit being reached prevents particle pair production occurring. Therefore, in contrast to the Coulomb case, no pairs are produced in the non-singular static  metric. For the Florides and Soffel metrics the singularity occurs in the black hole limit, while for the Schwarzschild interior metric it corresponds to infinite pressure at the centre. The behaviour of the energy level spectrum is discussed in the context of the semi-classical approximation and using general properties of the metric.
\end{abstract}

\section{Introduction}
\label{sec:intro}

In this work we consider the problem of a spin-half fermion gravitationally bound by a static, massive body described by a non-singular metric. We investigate the energy level spectrum of the particle in the limit that the metric exhibits a singularity. Previously, we calculated the spectrum of scalar (spin-zero) particles bound by such objects~\cite{Gossel2011} and found that it collapses to zero energy and becomes quasi-continuous in the singular limit. Importantly, bound states with zero energy (where the binding energy is equal in magnitude to the rest mass of the bound particle) were found not to exist when the metric is not singular.


Here we calculate the spin-1/2 fermion energy level spectrum for three interior metrics describing the interior of the gravitating (stationary) body: the Florides, Schwarzschild, and Soffel metrics, the latter so-named for its derivation and subsequent use in~\cite{Soffel1977}. For each metric we calculate the energy spectrum of the particle numerically and using approximate analytic expressions in the near-singular limit of the metric. In the Florides and Soffel spacetimes this limit is characterized by the approach of the Schwarzschild radius $r_s=2GM/c^2$ towards the radius of the body $R$, where $G$ is the gravitational constant and $M$ is the mass of the central body (black hole limit). In the case of the Schwarzschild interior metric the singularity occurs for $R=9r_s/8$ and corresponds to a infinite pressure at the origin, rather than the formation of an event horizon.

Our results show that, as in the scalar case, in the limit that the interior metric becomes singular, the ground state energy tends to zero; that is, the binding energy tends to the rest mass energy of the bound particle.  Additionally, the energy spectrum becomes quasi-continuous in this limit. This means that one cannot increase the strength of the central potential (by increasing $M$) to lower a bound state level from the positive continuum to the negative: one cannot reach the Dirac Sea to facilitate pair production. This suggests that pair-production cannot occur in the field of such a static metric prior to the formation of a black hole.

Additionally, we present a discussion of the semi-classical approach to analyzing the properties of the level spectrum. It is shown that as the metric develops a singularity, all the semi-classical levels will tend to zero. This behaviour can also be understood from classical quantities such as time dilation on the interior.

\section{Radial Wave Equations}
One may write a general static, spherically symmetric metric in the form (units $\hbar= c=1$)
\begin{equation}
\label{GeneralMetric}
ds^{2}= - e^{\nu \left( r \right)}dt^{2} + e^{\lambda \left( r \right)}dr^{2} +r^2 d\Omega^2 \,.
\end{equation}
The wavefunction for a spin-half particle (Dirac spinor) in a spherically symmetricl field may be separated using
\begin{equation}
\label{SepVarSubs}
\Psi = \frac{e^{-\frac{\nu}{4}}}{ (\sin{\theta})^{1/2}} R({r}) \Theta({\theta, \phi})e^{-i\epsilon t},
\end{equation}
with
\begin{gather}
\label{RTheta}
R({r}) =\frac{1}{r} \begin{pmatrix}
     f(r)    \\
     g(r)  
\end{pmatrix}\,,
\qquad
\Theta({\theta, \phi}) = \begin{pmatrix}
       \Omega_{\kappa\, m}   \\
       \Omega_{-\kappa\, m} 
\end{pmatrix}\,.
\end{gather}
Here $\kappa = (-1)^{l+j+1/2}(j+1/2)$ parametrises the angular momentum for a particle with well-defined values of orbital and total angular momentum, $j = l\pm 1/2$. The coupled radial wave equations are then (see, e.g.~\cite{Soffel1977,LandauLifshitz4} and Appendix A):
\begin{gather}
\label{CoupledRadialEqs}
\frac{df({r})}{dr} + e^{\frac{\lambda}{2}} \frac{\kappa}{r} f({r}) - e^{\frac{\lambda}{2}} \left[ e^{-\frac{\nu}{2}}\epsilon + m  \right] g({r}) = 0, \\
\label{CoupledRadialEqsB}
\frac{dg({r})}{dr} -e^{\frac{\lambda}{2}} \frac{\kappa}{r} g({r}) + e^{\frac{\lambda}{2}} \left[ e^{-\frac{\nu}{2}}\epsilon - m  \right] f({r}) = 0 \,.
\end{gather}

These equations can be decoupled to give a single second-order equation for the upper component $f(r)$
\begin{align}
\label{eq:SecondOrderEqn}
f''(r)& -\frac{1}{2}\left(\lambda' -\frac{\epsilon \nu'}{m e^{\nu/2}+\epsilon}\right)f'(r) \notag{} \\
+&\left[
 e^{\lambda}\left(\frac{\kappa^2}{r^2}-m^2+\epsilon^2 e^{-\nu}\right)+\kappa e^{\lambda/2}\left(\frac{\epsilon \nu'}{2m r e^{\nu/2}+2r\epsilon}+\frac{1}{r^2}\right)\right]f(r)=0.
\end{align} 
In order to determine the bound states of a spin-1/2 particle in the gravitational field of a finite-sized object we require a wave function that exists for all $r$. This necessitates an interior as well as an exterior metric to substitute into \eref{eq:SecondOrderEqn} and \eref{SepVarSubs}. The metric on the exterior of a spherically symmetric non-rotating body of mass $M$ is
\begin{equation}
\label{eq:ExteriorMetric}
e^\nu = \left(1-\frac{r_s}{r}\right) = e^{-\lambda}, r\geq R,
\end{equation}
where $r_s = 2GM$ is the Schwarzschild radius of the gravitating object. This, along with a suitable interior metric, allows us to construct interior and exterior wave equations of the form \eref{eq:SecondOrderEqn} which together span all $r$. These are solved both numerically and using approximate analytical formulae. 


\section{Method}
\subsection{Numeric Approach}
The bound states are found by numerically integrating interior and exterior versions of \eref{eq:SecondOrderEqn} using \textit{Mathematica} \cite{math} with the boundary condition for small $r$ determined by solving the zero order expansion of the interior wave equation around $r = 0$. The interior wave function is propagated to the boundary where continuity of the functions $f({R})$ and $f^{\prime}({R})$ is imposed. The solution is then propagated to large $r$ and the boundary condition $f(r\rightarrow \infty) = 0$ is enforced by varying $\epsilon$. The s-wave bound states for a given $R$ and $m$ are computed as a function of $r_s$. These results are compared with analytic calculations in later sections. 

\subsection{Harmonic Approximation}
\label{sec:HarmonicSection}
In the strong field limit (as the metric becomes singular) 
 the wave function of a bound particle is 
 contained entirely on the interior region of the body. The assumption of near complete containment on the interior allows one to perform all calculations using only the interior wave equation. The validity of this assumption is confirmed in subsequent sections. On the interior it is convenient to switch to dimensionless variables. We rescale the energy by the mass such that  $\varepsilon = \epsilon/m$, and rescale all lengths, including the Compton radius of the quantum particle, by the radius of the object $R$: $\rho=r/R$, $s = r_s/R$, $\mu = mR$.

One can re-write \eref{eq:SecondOrderEqn} in the form of a one dimensional Schr\"odinger equation using the transformation $f(\rho) = T(\rho) \psi(\rho)$ where
\begin{equation}
\label{eq:TransformationGen}
T(\rho) =\left(e^{\lambda/4}\sqrt{\mu(1+\varepsilon e^{-\nu/2})}\right)^{-1}
\end{equation}
yielding
\begin{equation}
\label{eq:GenEqn}
-\psi''(\rho)+V(\rho) \psi(\rho) = 0
\end{equation}
where $V(\rho)$ is an effective (energy dependent) potential. 

Analytic calculations of the bound state spectrum in the tight-binding limit can be performed by approximating the effective potential on the interior as that of a harmonic oscillator. For our interior metrics, $g^{\mu \nu}$, one may Taylor expand $V(\rho)$ near the origin and match the resulting expression to that of a quantum harmonic oscillator, given by
\begin{equation}
\label{eq:QHOV}
V_{\mathrm{HO}}(\rho) = \frac{l (l+1)}{\rho^2}+\mu^2\omega^2 \rho^2 -2 \mu E
\end{equation}
and use known solutions $E_n  = (n+3/2)\omega$, where $n = 2k + l$, $k \geq 0$, to approximate the energy levels of the system. Here $E_n = E_n (\varepsilon, s, \kappa, \mu)$ is the effective energy of the harmonic oscillator system. 

In the Florides case this approach is valid all the way to the singular metric limit. As will be seen in later sections, this is not true for the Soffel and Schwarzschild metrics where the harmonic approximation breaks down as the metric develops a singularity.

\subsection{Semi-classical approximation}
\label{sec:SCSection}
As the metric develops a singularity, the potential well becomes increasingly deep causing the wavefunction to oscillate many times on the interior. In this case one may invoke the semi-classical approximation and calculate excited energy levels using Bohr-Sommerfield quantization. When treating the problem semi-classically, all three singular metric limits may be taken.

To do this we investigate Eqn.~\eref{eq:GenEqn} for a metric possessing a singularity such that $e^{\nu(r)}\rightarrow 0$ for some value of $r$. In 
this limit Eqn.~\eref{eq:GenEqn} reduces to 
\begin{equation}
\label{eq:GenLimit}
-\psi''(\rho)+\psi(\rho)\left[\mu^2e^{\lambda}- \eps^2\mu^2 e^{(\lambda-\nu)}\right] = 0
\end{equation}
allowing us to define the approximate semi-classical momentum
\begin{equation}
\label{eq:SCMomentum}
p(\rho) =  \left[\eps^2\mu^2 e^{(\lambda-\nu)}-\mu^2e^{\lambda}\right]^{1/2}.
\end{equation}
Assuming that the semi-classical approximation is valid, the total phase accumulated at the boundary will contain the integral
\begin{align}
\label{eq:BohrSommerfield}
\int_{\rho_1}^{\rho_2} p(\rho) d\rho &\approx \int_{\rho_1}^{\rho_2}\left[\eps \mu e^{(\lambda-\nu)/2}-\frac{\mu e^{(\lambda+\nu)/2}}{2\eps}\right] d\rho  \notag{} \\
&=X\eps \mu-\frac{Y \mu}{\eps}
\end{align}
Setting this phase to $k \pi$ yields the energy levels as
\begin{equation}
\label{eq:GenEn}
 \eps_k = \frac{\pi k +\sqrt{\pi^2k^2+4\mu^2XY}}{2X\mu}.
\end{equation}
This expression tends to zero in the singular limit if $Y/X\rightarrow 0$, which is satisfied in this limit as $X$ contains the most singular term. Thus all levels tend to zero and the spectrum collapses.

Additionally, one may arrive at this conclusion by directly considering the metric. In the singular limit, the coefficient of $dt^2$ in the metric tends to zero giving rise to (local) time dilation that approaches infinity, with the classical period $T$ of a bound particle being similarly affected. Therefore in this limit the semi-classical level spacing becomes $\omega = 2\pi/T \rightarrow 0$ and we arrive at the previous result.

\section{Solution of the wave equation in the Florides interior metric}


An interior metric that does not contain the singularity present in the Schwarzschild interior solution, discussed in Section~\eref{sec:schwarzschild}, is that developed by Florides~\cite{Flor74}:
\begin{align}
\label{eq:InteriorMetric}
 e^{\nu} &=  \frac{\left(1-s\right)^{\frac{3}{2}}}{\sqrt{1-s \rho^2}},\notag{} \\ 
e^{\lambda} &= \left(1-s \rho^2\right)^{-1}
\end{align}
which is valid for $r\leqslant R$. For $r_s=R$ the horizon (coordinate) singularity exists at $r=r_s$~\cite{comment}. The Florides metric corresponds to a spherically symmetric, constant density body with vanishing radial stresses. 
For our purposes this metric is characterized by its effective potential $V(\rho)$ which, near $\rho = 0$ (for $s\rightarrow 1$), is given by 
\begin{align}
\label{eq:FloridesHOExp}
V(\rho) &\approx \frac{\kappa(\kappa + 1)}{\rho^{2}} + \left( \mu^{2} + s\kappa^{2} - \frac{s}{4} - \frac{\mu^{2}\varepsilon^{2}}{(1-s)^{3/2}}  \right) \\ \notag{}
 +& \left( \mu^{2}s + \kappa s^{2} \left( \kappa-\frac{3}{8}\right)  - \frac{11 s^{2}}{16} - \frac{\varepsilon^{2} \mu^{2} s}{2(1-s)^{3/2}} \right)\rho^{2} + O\left(\rho^4\right).
\end{align}
 Equating the coefficients of the powers of $\rho$ between Eqns.~\eref{eq:QHOV} and \eref{eq:FloridesHOExp}, we derive two equations for the two unknowns $\omega$ and $\varepsilon$ and can therefore solve for both. The energy solutions that corresponds to the energy eigenvalues are
\begin{equation}
\label{eq:FlorEpsilon}
\varepsilon = \pm \frac{(1-s)^{3/4}}{\mu}\sqrt{\alpha_\text{F} + \beta_\text{F}}
\end{equation}
where
\begin{align*}
\alpha_\text{F} &= \mu^{2} + s \left( \kappa^{2} - \frac{1}{4} \right) - s \left( n+\frac{3}{2} \right)^{2}, \\
\beta_\text{F} &=  s \left(n+\frac{3}{2}\right) \sqrt{\left(n+\frac{3}{2}\right)^{2} + \frac{2\mu^{2}}{s} + 2\kappa \left( \kappa - \frac{3}{4} \right)-\frac{9}{4}} .
\end{align*}

The above expressions are valid under the assumption that the entire wavefunction is contained within the interior region $\rho \leq 1$ ($r \leq R$). This condition can be satisfied when $\mu$ is large, forcing the particle to be bound near the origin. Formally we require
$\langle \rho^2 \rangle \ll 1$.
The value of $\langle \rho^2 \rangle$ can be obtained using the virial theorem for our harmonic potential
\begin{align}
\left< \frac{\mu \omega^2 \rho^2}{2} \right> = \frac{1}{2}(n+3/2)\,\omega \notag \\
\label{eq:BindingCondition}
\langle \rho^2 \rangle = \frac{(n+3/2)}{\mu\omega} \ll 1 .
\end{align}
Expanding $\omega$ about $1/\mu$  in \eref{eq:BindingCondition} using \eref{eq:FlorEpsilon} to lowest order gives $\omega \sim \sqrt{s/2}$ where our truncation of the expansion is valid for 
\begin{equation}
\label{eq:FC1}
\mu \gg \sqrt{\frac{s}{2}}\sqrt{\kappa^2+2k^2},
\end{equation}
recalling that $n = 2k+l$. Using this value of $\omega$ the condition \eref{eq:BindingCondition} becomes
\begin{gather}
\label{eq:InteriorCondition}
\mu \gg \sqrt{\frac{2}{s}}(n+3/2)\ 
\end{gather}
which automatically satisfies Eqn.~\eref{eq:FC1}.

A second test is whether the expansion of $V_I(\rho)$ as a harmonic potential is valid. The ratio of the fourth order to second order term (at the expected particle position) is, for large $\mu$, given by
\[
\left|\frac{\langle \rho^2 \rangle A_4}{A_2}\right| =\frac{5}{4\mu}(n+3/2)\sqrt{2s}
\]
where $A_n$ is the coefficient of the $\rho^n$ term in the expansion \eref{eq:FloridesHOExp}. Where the harmonic oscillator model is appropriate this should be much less than one, which leads to
\begin{equation}
\mu \gg\frac{5}{4}(n+3/2)\sqrt{2s}.
\end{equation}
This condition is automatically satisfied by Eqn.~\eref{eq:InteriorCondition}.
Furthermore, $\omega > 0$ is satisfied automatically (when \eref{eq:InteriorCondition} is satisfied) as it must to be physically valid.
Finally, we note that in the limit $\mu \gg n$, we can expand the energy \eref{eq:FlorEpsilon} as
\begin{equation}
\varepsilon = \pm (1-s)^{3/4}\left(1+(n+3/2)\sqrt{\frac{s}{2\mu^2}}+O\left(\frac{s}{\mu^2}\right)\right)\ .
\end{equation}
This is always $\varepsilon < 1$ under condition \eref{eq:InteriorCondition} (that is, the state is bound).

\begin{figure}[h!]
\begin{center}
\includegraphics[width=10cm]{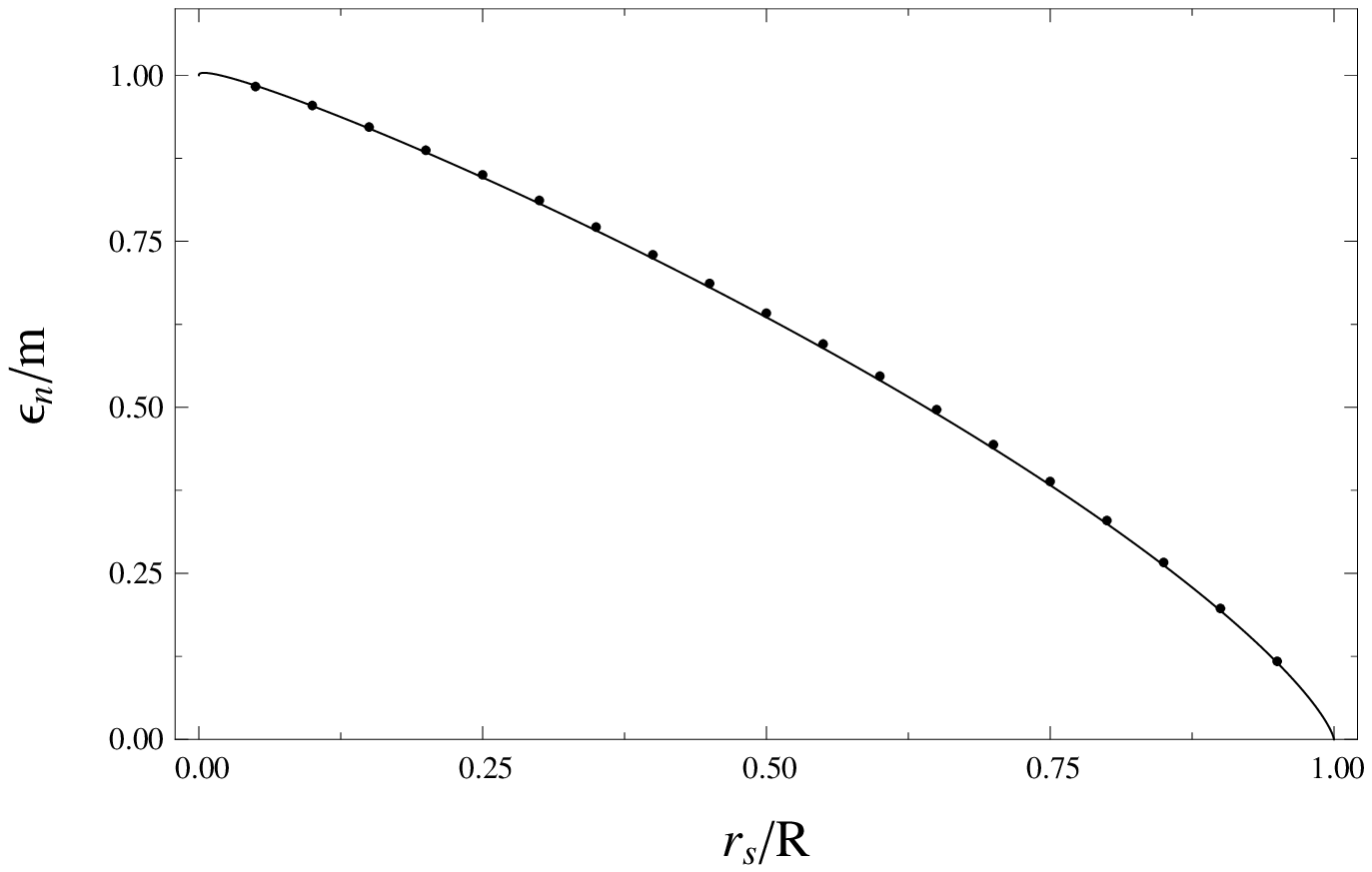}
\caption{Ground state energy of spin-1/2 particle in Florides space-time with $\mu=10$, $n=0$ and $\kappa =-1$. Circles: numeric; solid line: analytic approximation \eref{eq:FlorEpsilon}.}
\label{fig:FloridesEPlot}
\end{center}
\end{figure}

\begin{figure}[h!]
\begin{center}
\includegraphics[width=10cm]{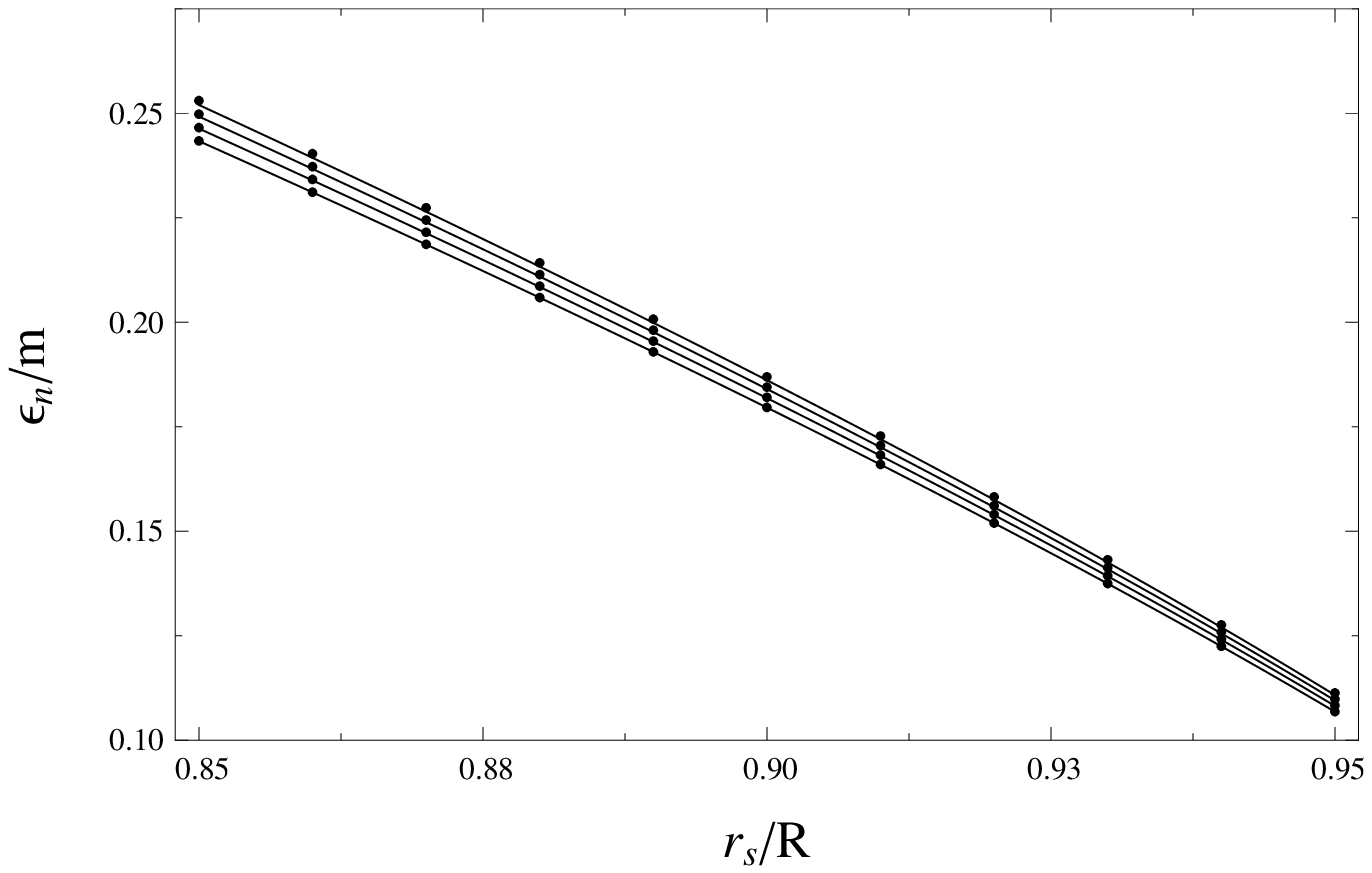}
\caption{Lowest 4 energy levels of spin-1/2 particle in Florides space-time with $\mu=100$, and $\kappa =-1$.  Circles: numeric; solid line: analytic approximation \eref{eq:FlorEpsilon}.}
\label{fig:FloridesExcitedPlot}
\end{center}
\end{figure}

Figure \eref{fig:FloridesEPlot} shows the numeric ground state energy as a function of $s$ compared with Eqn.~\eref{eq:FlorEpsilon}. Figure  \eref{fig:FloridesExcitedPlot} compares the analytics and numerics for the ground and first three excited states.
The figures show that there is good agreement between our numerical and analytic spectrum. A large mass of $\mu = 100$ was chosen for Fig.~\eref{fig:FloridesExcitedPlot} in order to satisfy the condition (\ref{eq:InteriorCondition}), which is better fulfilled as $\mu$ increases, ensuring the validity of the harmonic approximation for all $n$ displayed.



\section{Solution of the wave equation in the Soffel interior metric}

As an alternative to the approach adopted by Florides, Soffel \textit{et al.} extended the interior Schwarzschild metric past the pressure singularity by analytic continuation \cite{Soffel1977}. 
We perform our analysis of the bound state spectrum using the same procedures and techniques employed in the previous two sections. 
 This `Soffel' metric corresponds to
\begin{align}
\label{SoffelMetric}
e^\nu &= (1-s) \exp\left[- \frac{s (1-\rho^2)}{2 (1-s)}\right], \\
e^\lambda &= \left(1-s \rho^2\right)^{-1}\notag{}.\notag{} \\
\end{align}
Using the methods outlined previously, we find in the Soffel metric the scaled energy, $\varepsilon$, is
\begin{equation}
\label{eq:MainSoffelEnergy}
\varepsilon = \pm \frac{(1-s)^{1/2}\exp[-\frac{s}{4(1-s)}]}{\mu}\sqrt{\tau + \chi},
\end{equation}
where
\begin{align}
\label{MainSoffelTau}
\tau &= \mu^2+s\left(\kappa^2 +\frac{(\kappa-1)(1-2s)}{4(1-s)} - \frac{(1-2s)(n+3/2)^2}{1-s}\right), \notag{} \\
\chi &=  \sqrt{\tau^{2} + 4C\left( n+3/2 \right)^{2} -\left(\tau+\frac{s (1-2s)(n+3/2)^2}{1-s}\right)^2}, \notag{}\\
C&=s \left(\mu^2-\frac{9\kappa s}{8}+s\kappa^2+\frac{3}{4}-\frac{3}{8 (1-s)}\right)\notag{} \\
&+ \left(\frac{s}{4(1-s)}\right)^2 \left(4\left[2s^2 (3\kappa-4)+\kappa (5-11s)+14s\right]-26\right).
\end{align}

\begin{figure}[h!]
\begin{center}
\includegraphics[width=10cm]{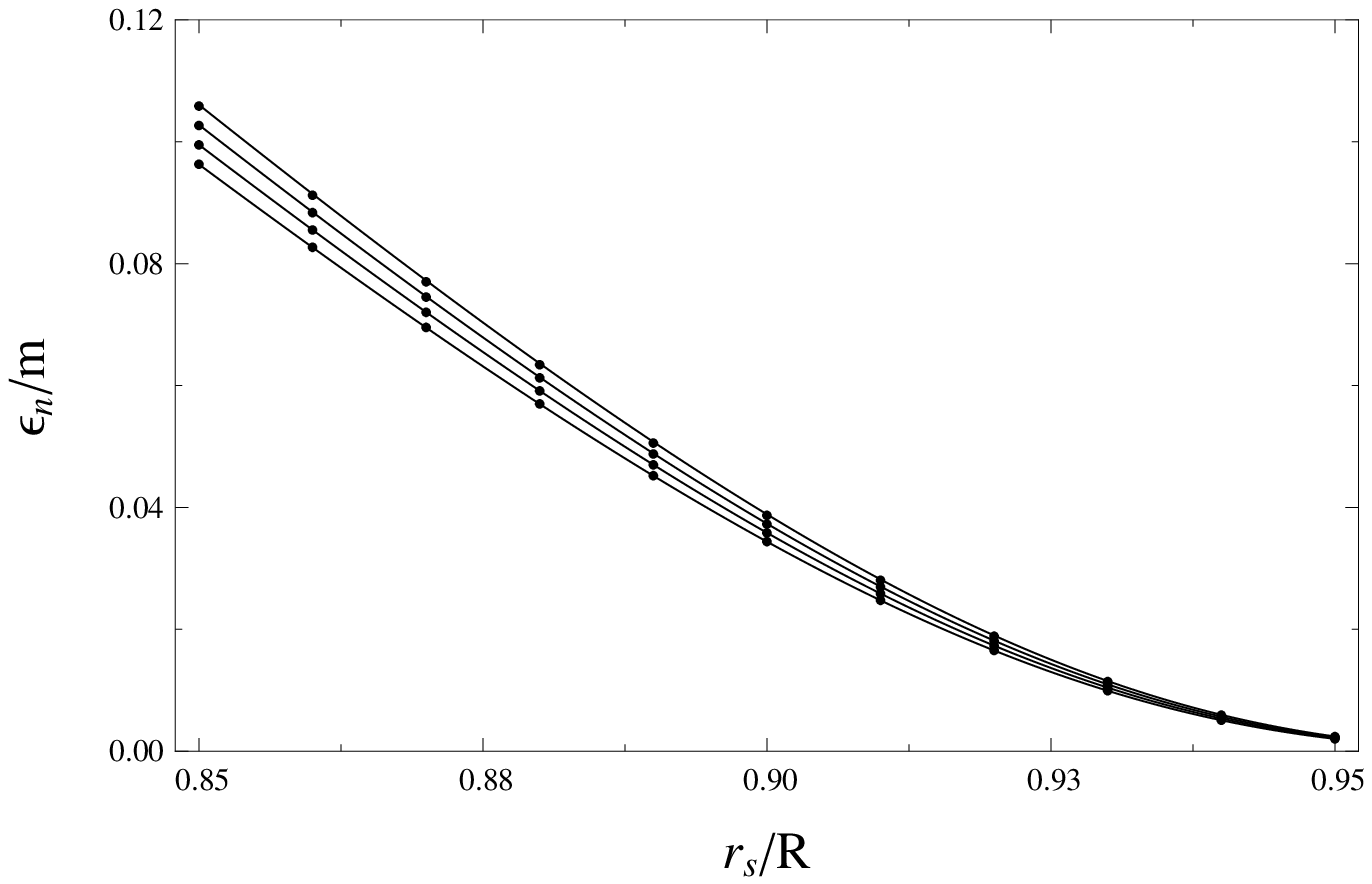}
\caption{Lowest 4 energy levels of spin-1/2 particle in Soffel space-time with $\mu=100$, and $\kappa =-1$. Circles: numeric; solid line: analytic approximation \eref{eq:MainSoffelEnergy}.}
\label{fig:SoffelExcitedPlot}
\end{center}
\end{figure}

As in the previous case we examine conditions on $\mu$ (taken to be large) such that $\langle \rho^2\rangle\ll1$ and the quadratic term in the expansion of the potential dominates (using Eqn.\eref{eq:MainSoffelEnergy} as $\varepsilon$). This yields the condition
\begin{equation}
\label{eq:SoffelCond1}
\mu \gg \frac{\sqrt{1-s}(n+3/2)}{2\sqrt{2s}},
\end{equation}
which is automatically satisified for $s\rightarrow 1$ (a condition on $\mu$ w.r.t. $\kappa$ and $k$ can be generated in a similar fashion to Eqn.~\eref{eq:InteriorCondition} of the Florides case, but is automatically satisfied by \eref{eq:SoffelCond1}).
 This is due to the presence of extra $1/(1-s)$ terms in the expansion of the effective potential corresponding to the Soffel metric. This also affects the ratio of the fourth order to second order coefficients. We find that, unlike in the Florides case, the effective potential changes in shape significantly near the origin as $s\rightarrow 1$. Thus, in order to ensure that the potential where the particle is bound remains harmonic we require a more stringent condition on $\mu$. Specifically, we find that
\begin{align}
\label{eq:SoffelMassCondition}
\left|\frac{\langle \rho^2 \rangle A_4}{A_2}\right|  &= \frac{(n+3/2) \sqrt{s}}{2\mu\sqrt{2}\sqrt{1-s}}\ll 1, \notag{} \\
\implies \mu &\gg \frac{(n+3/2) \sqrt{s}}{2\sqrt{2}\sqrt{1-s}}.
\end{align}
For comparison with results presented in Fig.~\eref{fig:SoffelExcitedPlot} for $n = 4, s=0.95$ this corresponds to $\mu\gg 8$.

\begin{figure}[h!]
\begin{center}
\includegraphics[width=10cm]{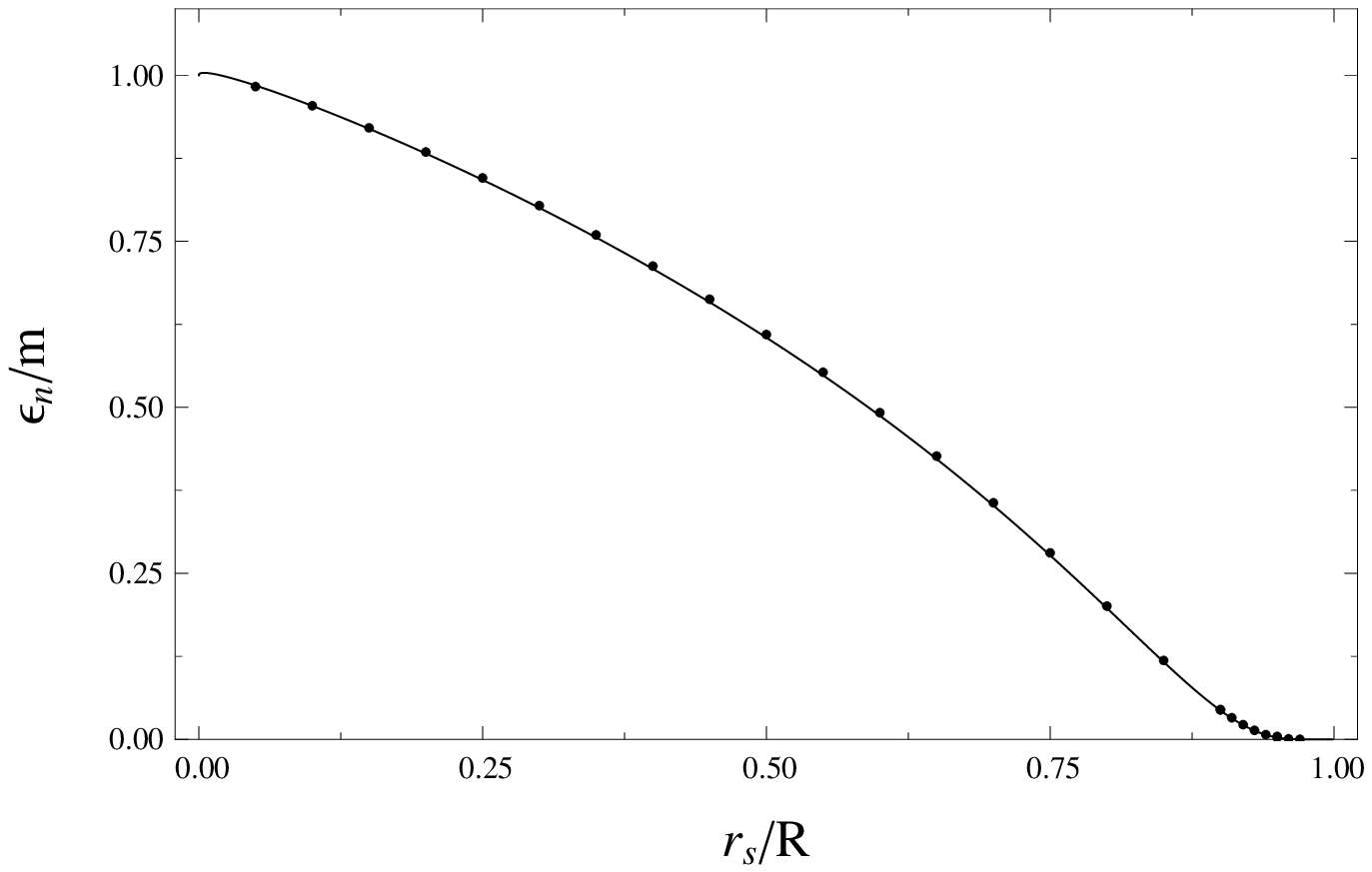}
\caption{Ground state energy of spin-1/2 particle in Soffel space-time with $\mu=10$, $n=0$ and $\kappa =-1$. Circles: numeric; solid line: analytic approximation \eref{eq:MainSoffelEnergy}.}
\label{fig:SoffelEPlot}
\end{center}
\end{figure}

As $s\rightarrow 1$ the energy expression (\ref{eq:MainSoffelEnergy}) 
becomes
\begin{equation}
\label{ReducedSoffelEnergy}
\varepsilon \approx \pm \exp\left[-\frac{s}{4(1-s)}\right]\sqrt{1-s}\left(1+\frac{(n+3/2)\sqrt{s}}{\sqrt{2}\mu\sqrt{1-s}}\right)
\end{equation} 
which tends to zero in the black hole limit for both positive and negative energy states (Eqn.~\eref{eq:SoffelMassCondition} ensures the second term is $\ll 1$ in this limit).
Figures~\eref{fig:SoffelEPlot} and \eref{fig:SoffelExcitedPlot} show the ground state and excited state spectra respectively, with the numeric and  analytic results displayed with each other as before.
In the singular limit, $s\to1$, the spectrum (\ref{eq:MainSoffelEnergy}) is observed to exhibit the same limit as the Florides spectrum, (\ref{eq:FlorEpsilon}). The positive energy spectrum collapses to zero from above as the gravitating body tends towards the black hole limit, and the negative energy levels (Dirac sea) also collapse to zero. 

As Eqn.~\eref{eq:SoffelMassCondition} indicates, and as was noted in Sec.~\eref{sec:HarmonicSection}, the harmonic approximation breaks down in the singular limit for this metric, i.e. only in the limit of infinite $\mu$ is the harmonic approximation valid. Using the definitions in Eqn.~\eref{eq:BohrSommerfield} we calculate 
\begin{align}
X &\stackrel{s\rightarrow1}{=} \sqrt{\pi} \exp\left[\frac{1}{4(1-s)}\right], \notag{} \\
Y &\stackrel{s\rightarrow1}{=}0
\end{align}
yielding $\eps \rightarrow 0$ for all levels as $s\rightarrow 1$.

\section{Solution of the wave equation in the Schwarzschild interior metric}
\label{sec:schwarzschild}
As previously noted, the standard Schwarzschild interior metric cannot be used to investigate the limit $s\rightarrow 1$. This is due to a pressure singularity forming at the origin for $s=\nicefrac{8}{9}$ where the coefficient of $dt^2$ is zero.

 The interior solution for a constant density fluid sphere developed by Schwarzschild~\cite{Schwarzschild} is given by
\begin{align}
\label{eq:SchwarzschildInterior}
e^\nu &=\left(\frac{3}{2}\sqrt{1-s}-\frac{1}{2}\sqrt{1-s \rho^2}\right)^2, \notag{} \\
e^\lambda &= \left(1-s \rho^2\right)^{-1}.
\end{align}

Taking the singular limit ($s \rightarrow 8/9$) of the effective potential yields
\begin{multline}
\label{MainSchwarzExpansion}
V(\rho) = \frac{\kappa(\kappa+1)}{\rho^{2}} + \left( \mu^{2} + s\kappa^{2} -\frac{s\kappa}{2} +\frac{s(1-2\kappa)}{2(3\sqrt{1-s}-1)} - \frac{4\mu^{2}\epsilon^{2}}{(3\sqrt{1-s}-1)^{2}}   \right) \\
 + s \left( \mu^{2} - \frac{9s\kappa}{8} +s\kappa^{2} + \frac{s(\kappa-1)}{2(3\sqrt{1-s}-1)^{2}} + \frac{4(2-3\sqrt{1-s})\mu^{2}\epsilon^{2}}{(3\sqrt{1-s}-1)^{3}} \right)\rho^{2} + O(\rho^{4}).
\end{multline}
By equating powers of $\rho$ between (\ref{eq:QHOV}) and (\ref{MainSchwarzExpansion}) one can solve for $\varepsilon$, giving the energies as
\begin{equation}
\label{eq:MainSchwarzEnergy}
\varepsilon = \pm \frac{(3\sqrt{1-s}-1)}{\mu}\sqrt{\alpha_\text{S} + \beta_\text{S}},
\end{equation}
where
\begin{align}
\label{eq:MainSchwarzDelta}
\alpha_\text{S} &=  \frac{s(2-3\sqrt{1-s})(n+3/2)^{2}}{2(3\sqrt{1-s}-1)} +\delta,\\
\label{eq:MainSchwarzSigma}
\beta_\text{S} &=  \sqrt{\alpha_\text{Sc}^{2} - \delta^{2} + \frac{s(n+3/2)^{2}\gamma}{4}  },\\
\delta &=\frac{1}{4}\left(\mu^2+s \kappa \left(\kappa-\frac{1}{2}\right)\right),\\
\gamma &= \mu^{2}+  s\kappa\left(\kappa-\frac{9}{8} \right) + \frac{s\left(\kappa-1\right)}{2\left(3\sqrt{1-s}-1\right)^{2}}.
\end{align}
 
We can now check the validity of the approximations used. 
Once again invoking the virial theorem and using Eqn.~\eref{eq:BindingCondition}, the condition $\langle \rho^{2} \rangle \ll 1$  is satisfied for
\begin{equation}
\label{eq:SCond2}
\mu\gg \frac{(n+3/2)\sqrt{3\sqrt{1-s}-1}}{m\sqrt{s}}  \stackrel{s \rightarrow 8/9}{=} \frac{9 (n+3/2)\sqrt{8/9-s}}{4}.
\end{equation}
As with Eqn.~\eref{eq:InteriorCondition} in the Florides case, a condition on $\mu$ w.r.t. $\kappa$ and $k$ can be generated, but is automatically satisfied by \eref{eq:SCond2}. As done previously we require the quadratic terms in the expansion to dominate over higher order terms. In this case the Schwarzschild metric resembles the Soffel metric considered previously, as it too has an additional condition on $\mu$ compared to the Florides metric. Specifically, we find:
\begin{align}
\left|\frac{\langle \rho^2 \rangle A_4}{A_2}\right| &=\frac{(n+3/2)}{6\mu\sqrt{8/9-s}}\ll 1, \notag{} \\
\implies \mu &\gg \frac{(n+3/2)}{6\sqrt{8/9-s}}.
\end{align}
By taking into account condition \eref{eq:SCond2} we simplify the energy expression \eref{eq:MainSchwarzEnergy} to
\begin{equation}
\label{ReducedSchwarzEnergy}
\varepsilon \approx \pm \frac{3\sqrt{1-s}-1}{2}\left(1+\frac{(n+3/2)\sqrt{s}}{\mu \sqrt{3\sqrt{1-s}-1}}\right)
\end{equation}
which retains the behaviour $\lim_{s \to 8/9}{\varepsilon} = 0$ for both positive and negative energy states. 

\begin{figure}[h!]
\begin{center}
\includegraphics[width=10cm]{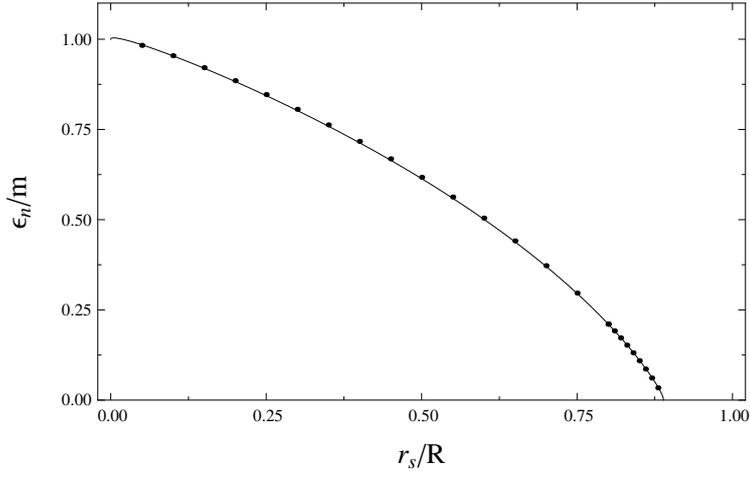}
\caption{Ground state energy of spin-1/2 particle in Schwarzschild interior space-time with $\mu=10$, $n=0$ and $\kappa =-1$. The metric becomes singular as $s$ approaches $8/9$ causing the binding energy to tend towards the rest mass of the fermion.  Circles: numeric; solid line: analytic approximation \eref{eq:MainSchwarzEnergy}.}
\label{fig:SchwarzEPlot}
\end{center}
\end{figure}
\begin{figure}[h!]
\begin{center}
\includegraphics[width=10cm]{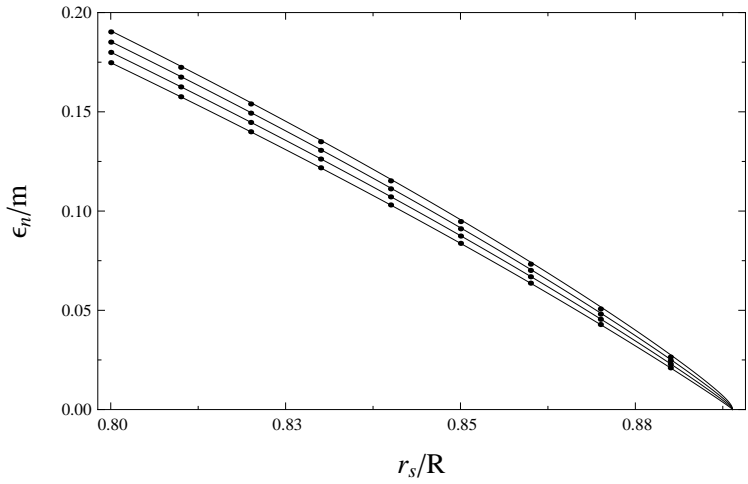}
\caption{Lowest 4 energy levels of spin-1/2 particle in Schwarzschild interior space-time with $\mu=100$, and $\kappa =-1$.  Circles: numeric; solid line: analytic approximation \eref{eq:MainSchwarzEnergy}.}
\label{fig:SchwarzExcitedPlot}
\end{center}
\end{figure}

Figures \eref{fig:SchwarzEPlot} and \eref{fig:SchwarzExcitedPlot} show the ground state and excited state spectra respectively, with the numeric and  analytic results displayed with each other as before.
As in the Florides case, our analytic formula reproduces the bound state energies well. In the singular limit, the spectrum is observed to exhibit the same qualitative behaviour as that in the Florides case. The positive energy spectrum collapses to zero from above, but with one important difference: the gravitating body develops a pressure singularity, causing the spectrum to collapse, well before the back hole limit. 

As in the Soffel case, the harmonic approximation breaks down when attempting to model the Schwarzschild case in the $s\rightarrow 8/9$ limit; one must take the infinite $\mu$ limit to ensure applicability as per Eqn.~\eref{eq:SCond2}. The corresponding calculation in the semi-classical picture is done using 
\begin{align}
X &\stackrel{s\rightarrow8/9}{=}\frac{\pi}{\sqrt{2(8/9-s)}}, \notag{} \\
Y &\stackrel{s\rightarrow8/9}{\approx}0.7.
\end{align}
Substituting this into Eqn.~\eref{eq:GenEn} yields $\eps \rightarrow 0$ for all levels as $s\rightarrow 8/9$.

\section{Discussion}
\label{sec:discussion}
We present calculations of the energy level spectrum for a massive spin-1/2 particle bound by a spherical static body. This is done by matching a suitable interior metric to the standard Schwarzschild exterior metric and solving the Dirac equation. 
Numerical computation of the ground state energies of the particle shows that the existence of a zero energy bound state may only be possible in the limit that the interior metric develops a singularity. In the case of the Florides and Soffel metrics this corresponds to the black hole limit; for the Schwarzschild interior metric this singularity represents infinite pressure at the centre. 
This result is verified using an approximate analytical calculation. For the Florides and Soffel metrics it is seen that the entire bound state spectrum collapses to $\epsilon = 0$ as $r_s\rightarrow R$: the energies and the intervals between the energy levels are proportional to $(1-r_s/R)^{3/4}$ and $\exp[-\frac{1}{(1-r_s/R)}]$ respectively. If we keep energy fixed the principal quantum number of the level $n$ tends to infinity. In the case of the Schwarzschild interior metric the singularity occurs for $r_s \rightarrow 8R/9$ with the spectrum going as $\epsilon \propto \sqrt{8/9-r_s/R}$.

 What is clear is that for any fixed $m$, $n$, $\kappa$, the general form of the equations for $\epsilon_n$ in the Florides and Soffel cases
ensures that the bound state energy collapses to zero in the limit $r_s \to R$. 
   Additionally, the form of the energy spectrum in the Schwarzschild interior case 
 results in the same collapse of the spectrum, except now the collapse occurs in the limit $r_s \to 8R/9$. This means that for spin-1/2 particles, as in the case of scalar particles, the event horizon singularity is not the only singularity that results in collapse of the spectrum to a gapless state --- a pressure singularity gives the same result. Therefore the bound state energy spectrum of an object with a near-singular metric reproduce those of an object that is close to the black hole limit.

The existence of a bound state with zero energy is relevant to the phenomenon of particle pair production. Due to quantum fluctuations, particle anti-particle pairs are produced around a black hole and it is possible that one of the pair escapes the gravitational field yielding Hawking radiation~\cite{Hawking74,Hawking75}.

Other systems that give rise to pair production, also known as vacuum breakdown, include static and dynamic electromagnetic fields (see, e.g.~\cite{Ruffini08}) and time-varying gravitational fields (see, e.g.~\cite{Hossenfelder}). In the Coulomb case it is clear that one of the pair must be repelled, since the pair have opposite charge. In the gravitational case, however, both particles will possess the same `gravitational charge' and thus will be attracted by the potential. In the black hole case the existence of the event horizon as a barrier facilitates one particle escaping to infinity as the other falls to the singularity.

For gapless energy states to exist in the metrics we have considered, the metric must become singular. For pair production with the ejection of one particle to infinity it is actually necessary to have a negative energy level $\epsilon < -mc^2$. As a concrete example, compare the gravitational field case with the electrostatic potential case where pair creation is possible in a strong Coulomb field $U(r)$. (For example, the ground state of an electron orbiting a finite-size nucleus reaches the lower continuum when $Z \gtrsim 170$ \cite{popov71sjnp}). In the Coulomb case we have  $(\epsilon - U(r))^2$ in the wave equation. Therefore, increasing the absolute value $|U|$ of the  negative potential $U(r)$ leads to the negative energy $\epsilon$. In the gravitational field we have the field dependent term $e^{-\frac{\nu}{2}}$ appearing as a multiplicative factor of the energy, rather than as a subtractive function, a situation that cannot introduce negative energy bound states.\\

\begin{acknowledgements}

We thank M. Yu. Kuchiev and G. F. Gribakin for useful discussions. This work is supported by the Australian Research Council.

\end{acknowledgements}

\appendix
\section{Derivation of Curved Space Dirac Equation}


Although the following discussion is already well represented in the literature (see, for example \cite{ParkerToms}), we restate the most important results here as a reference for the reader.

The Dirac equation on a curved manifold (with $\hbar =c=1$) reads
\begin{equation}
\label{CurvedDiracEq}
(i\tilde{\gamma}^{\mu}\nabla_{\mu} - m )\Psi = 0.
\end{equation}
The Dirac matrices $\tilde{\gamma}^{\mu}$ satisfy the Clifford algebra
\begin{equation}
\label{CurvedDiracAlgebra}
\tilde{\gamma}^{\mu} \tilde{\gamma}^{\nu} + \tilde{\gamma}^{\nu} \tilde{\gamma}^{\mu}  = 2 g^{\mu \nu},
\end{equation}
where $g^{\mu\nu}$ is the reciprocal of the metric and $m$ is the rest mass of the spin-1/2 particle. A tilde above a gamma matrix indicates that the matrix is to be taken in curved space-time, conversely, the absence of a tilde indicates that the gamma matrix is to be taken in flat Minkowski space-time. The covariant derivative is 
\begin{equation}
\label{CovDeriv}
\nabla_{\mu} \Psi = (\partial_{\mu} - \Gamma_{\mu})\Psi,
\end{equation} 
where the spin-affine connections, $\Gamma_{\mu}$, may be determined from the relation
\begin{equation}
\label{SpinConnection}
\Gamma_{\mu}(x)=-\frac{1}{4}\gamma_{\alpha} \gamma_{\beta} \eta^{\alpha \delta} b_{\delta}^{\ \lambda} \left( \partial_{\mu}b^{\beta}_{\ \lambda} - \Gamma^{\sigma}_{\ \mu \lambda} b^{\beta}_{\ \sigma} \right).
\end{equation}
The vierbein of vector fields, $b^{\alpha}_{\ \mu}$, are found form the equality
\begin{equation}
\label{VierbeinDefn}
g_{\mu \nu} = \eta_{\alpha \beta} b^{\alpha}_{\ \mu} b^{\beta}_{\ \nu},
\end{equation}
where $\eta_{\alpha \beta}$ is the Minkowski metric (with signature $(-+++)$ here). The Christoffel symbols $\Gamma^{\sigma}_{\ \mu \lambda}$ are \begin{equation}
\label{Christoffel}
\Gamma^{\sigma}_{\ \mu \lambda}(x) = \frac{1}{2}g^{\sigma \alpha} \left( g_{\alpha \mu , \lambda} + g_{\alpha \lambda , \mu} - g_{\mu \lambda , \alpha} \right).
\end{equation}
This then yields the equations for $f(r)$ and $g(r)$:
\begin{gather}
\frac{df({r})}{dr} + e^{\frac{\lambda}{2}} \frac{\kappa}{r} f({r}) - e^{\frac{\lambda}{2}} \left[ e^{-\frac{\nu}{2}}\epsilon + m  \right] g({r}) = 0, \\
\frac{dg({r})}{dr} - e^{\frac{\lambda}{2}} \frac{\kappa}{r} g({r}) + e^{\frac{\lambda}{2}} \left[ e^{-\frac{\nu}{2}}\epsilon - m  \right] f({r}) = 0.
\end{gather}
\end{document}